\documentstyle[epsf,amstex,onecolumn]{mn.a4}    
\def\lesssim{\mathrel{\hbox{\rlap{\hbox{\lower4pt\hbox{$\sim$}}}\hbox{$<$}}}}
\def\gtrsim{\mathrel{\hbox{\rlap{\hbox{\lower4pt\hbox{$\sim$}}}\hbox{$>$}}}}
\newcommand{\ltaraw}{$\; \buildrel < \over \sim \;$}
\newcommand{\lta}{\lower.5ex\hbox{\ltaraw}}
\newcommand{\gtaraw}{$\; \buildrel > \over \sim \;$}
\newcommand{\gta}{\lower.5ex\hbox{\gtaraw}}
\def\frac#1/#2{\leavevmode\kern.1em\raise.6ex\hbox{\the\scriptfont0 #1} 
            \kern-.15em/\kern-.2em\lower.7ex\hbox{\the\scriptfont0 #2}} 

\newcommand{\ie}{{\it i.e.~}}
\newcommand{\eg}{{\it e.g.~}}
\newcommand{\etal}{{\it et.~al.~}}
\newcommand{\cf}{{\it c.f.~}}

\newcommand{\kev}{{\rm\,keV}}
\newcommand{\msun}{{\rm\,M_\odot}}

\newcommand{\ergsec}{{\rm ergs\,s^{-1}}}
\newcommand{\ergs}{{\rm ergs}}
\newcommand{\cm}{{\rm\,cm}}

\newcommand{\g}{{\rm\,g}}

\newcommand{\zform}{{z_{\rm form}}}

\loadboldmathitalic 
\title[X-Ray Properties of Groups and Clusters of Galaxies]
{Physical Implications of the X-ray Properties of Galaxy Groups and Clusters}
\author[A. Babul \etal]
{Arif Babul$^{1}$, Michael L. Balogh$^{2}$, Geraint F. Lewis$^{3}$, Gregory B. Poole$^{1}$\\
$^{1}$ Department of Physics \& Astronomy, University of Victoria, Victoria, BC, V8P 1A1, Canada\\
Email: {\tt babul@@uvic.ca},\ \  {\tt gbpoole@@uvastro.phys.uvic.ca} \\
$^{2}$ Department of Physics, University of Durham, South Road, Durham, DH1 3LE, UK\\
Email: {\tt M.L.Balogh@@durham.ac.uk} \\
$^{3}$ Anglo-Australian Observatory,  PO Box 296, Epping, NSW 1710, Australia \\
Email: {\tt gfl@@aaoepp.aao.gov.au} \\
}

\date{\today}
\begin{document}

\maketitle 
\begin{abstract}
Within the  standard framework of structure  formation, where clusters
and  groups of  galaxies  are built  up  from the  merging of  smaller
systems, the  physical properties of the intracluster  medium, such as
the gas temperature  and the total X-ray luminosity,  are predicted to
possess well defined self-similar scaling relations. 
Observed clusters and groups, however, show strong deviations
from these predicted relations. We argue that these deviations 
are unlikely to be entirely due to observational biasses; we 
assume they are physically based, due to the presence of excess 
entropy in the intracluster medium in addition to that generated  
by  accretion  shocks  during the  formation  of  the cluster. 
Several  mechanisms  have  been  suggested as  a  means  of
generating this  entropy. Focussing  on those mechanisms  that preheat
the  gas before  it becomes  a constituent  of the  virialized cluster
environment,  we present  a simple,  intuitive,  physically motivated,
analytic model that successfully  captures the  important physics
associated  with the  accretion of  high  entropy gas  onto group  and
cluster-scale  systems.   We  use  the  model to  derive  the  new
relationships between the observable properties of clusters and groups
of  galaxies, as  well as  the  evolution of  these relations.   These
include the  luminosity-temperature and luminosity-$\sigma$ relations,
as well as the  temperature distribution function and X-ray luminosity
function.   These   properties  are  found  to  be   a  more  accurate
description of the observations than those predicted from the standard
framework.  Future observations that will further test the efficacy of
the preheated gas scenario are also discussed.
\end{abstract}
\section{Introduction}\label{sec-intro} 

In currently favoured  models for structure formation, virialized 
structures, such as groups and clusters of galaxies, are the results of 
sequences of gravitationally-driven accretion and mergers of 
smaller  ``building blocks''.  According to the simplest (and what has 
effectively come to be regarded as the ``standard'') of such 
models, the thermodynamic properties of the intracluster medium (ICM) 
are established by shocks and compression occuring during the accretion 
process. This model predicts that the intrinsic properties of the haloes 
(\eg~ mass, temperature, circular velocity, etc.) 
obey self-similar scalings. In turn, these imply that observable properties,
such as the X-ray luminosity (L) and the temperature (T) of 
the intracluster gas in clusters and groups ought to scale as
$L \propto T^2$ \cite{K91,ECF} and $L \propto T$ 
(Balogh, Babul \& Patton 1999), respectively.  Such predictions,
however, do not match the observed relation for either the clusters, 
which is approximately $L \propto T^3$ \cite{ES91,M98}, or the 
groups, which is argued to be even steeper \cite{P+96,HP00}.
The theoretical models are also unable to
account for the large, flat cores in the X-ray surface brightness profiles 
of the majority of the clusters \cite{L+00}.  Moreover, there is
an indication of a discrepancy between the theoretical and the observed
mass-temperature relationship.  
Several authors \cite{EF99,N00,XJW}
have fit the mass-temperature data for systems
with temperatures greater than 1 keV with a simple power-law and find
$M\propto T^{1.7-1.9}$, which is steeper than the standard 
theoretical result: $M\propto T^{3/2}$. Finally, the 
standard model also predicts an entropy-temperature relationship that appears
to be in conflict with the observed trend of a gradual
flattening towards low temperatures \cite{PCN,LPC00}.

Broadly speaking, proposed resolutions to these problems can be grouped into
three classes:  those which appeal to observational biases or neglected
physics such as cooling; those in which thermal energy is injected into the gas, either
before or after collapse; and models which invoke non-thermal processes.
We review these solutions in more detail below.

We start by addressing the possibility that no heating whatsoever is required.
Firstly, it is possible that many, if not most, of the discrepant trends
mentioned above are the result of  observational biases arising due to 
the low surface brightnesses of the lower mass systems \cite{M00};
we will discuss this issue in \S\ref{sec-discuss}.  
However,
the best way to approach this problem is to create ``mock'' images from the
models, and subject them to the same selection and reduction procedures as
the real data (Poole \etal, in preparation).  Secondly, most of the analytic models
of the ICM that we discuss below disregard
the effects of cooling.  Recent simulations, which include cooling 
(\eg Lewis \etal 2000; Pearce \etal 2000; Muanwong \etal 2001, Bryan \& Voit 2001), 
have shown that this can have a significant effect on cluster and group gas 
profiles. 
One possibility is that as the low entropy gas in the 
central regions cools, condenses into dense, cold structures,  and the higher entropy gas 
flows in to fill the volume, the X-ray luminosity will decrease.  
Whether this actually happens, however, 
has yet to be convincingly demonstrated.  Lewis \etal (2000) find that in 
their high resolution cluster simulations, the inclusion of cooling (and 
star formation) in fact leads to an increased X-ray luminosity.  Finally, 
Bryan (2000) has suggested that galaxy formation
is more efficient in groups than in clusters, which reduces the amount of hot
gas and therefore the X-ray luminosity.  However, this interpretation has been
challenged by Balogh \etal (2001), who claim that the observed trend on which
Bryan's conclusion is based is the result of biases in measuring the gas
fractions of galaxy groups.

We now turn our attention to models which use energy injection to break the
self-similar relations.  Such a model was originally proposed by Kaiser 
(1991), and has been subsequently explored by many authors 
(\eg Evrard \& Henry 1991; Bower 1997; Cavaliere \etal 1997, 1998; 1999;
Balogh, Babul \& Patton 1999; Wu, Fabian \& Nulsen 2000; Lowenstein 2000; 
Tozzi \& Norman 2001; Voit \& Bryan 2001).
In general, these models require that 
at least the central $\sim 10^{13} \msun$ of the ICM has been 
injected with energy unassociated with the collapse and virialization of the 
groups and clusters at the level of 1--3 keV per particle.

There are many mechanisms for injecting the required energy into the 
intracluster medium.  The one most commonly invoked is thermal energy from
supernova explosions 
occuring within the  group/cluster galaxies \cite{VS99,PCN,L00,BBBCLF,BM01}.  The studies that have considered
this possibility have, however, generally tended to conclude that even if the
heating were to take place before collapse (when the energetics are most
favourable), in order to have the required impact, the mean efficiency with 
which
the supernovae energy is deposited into the ICM must be very high {\it and}
that the
initial mass function in the galaxies must be skewed towards high masses
\cite{BBP99,VS99,L00,BBBCLF,BM01}.
The former is possible if a large fraction of the stars in the 
galaxies formed in intense starbursts \cite{H2000} or, as suggested by 
recent high-resolution simulations by Lewis \etal (2000), a significant 
fraction of the stars in the cluster are diffusely distributed within the 
intracluster volume. On the other hand,  there is no compelling evidence 
for an IMF that is very different from that observed locally (\eg Wyse 1997).  

There is potentially another large source of thermal energy in quasars and 
active galactic nucleii (\eg Valageas \& Silk 1999; Wu \etal~2000).  
Kormendy \& Richstone (1995) and  Magorrian \etal~(1998) present 
strong evidence that most spheroidal galaxies are likely to  
harbour massive black holes of typical mass $\sim 10^8\msun$ in their centers.
When fueled, these black holes are expected to behave like quasars and
under reasonable assumptions, there appears to be more than enough energy 
available from these objects to heat $\sim 10^{13} \msun$ of gas in
the surrounding environment to a temperature of $\sim 10^6$ K 
(\eg Fabian 1999).  Moreover, in order to account for the observed 
number density of quasars at $z\gta 2$, Silk \& Rees (1998) suggest 
that most of these black holes must have been active at some earlier 
epoch. The main drawback of this scheme is that the coupling between quasars
and their surrounding medium is not well understood.  One possibility
is that the ambient medium is heated by shocks and turbulence created 
by fast moving jets produced by the active black holes (c.f. Kaiser 
\& Alexander 1999; Rizza et al 2000).   Another possibility is that quasars 
eject nearly spherical high velocity outflows that then shock-heats 
the ambient intergalactic medium.  There is evidence for such
outflows: Studies of the UV Broad Absorption Lines (BALs) suggest 
that these are created in outflows with velocities reaching $\sim 0.1c$
and having covering factors as large as 50\% (\eg Brandt \etal 1999).

Yet another possible origin of pre-collapse heating is suggested by 
recent detailed high-resolution numerical studies of cluster 
formation in its proper cosmological setting  \cite{L+00} and 
of the evolution of the intergalactic medium \cite{C+95,CO99,D+00}.  
These simulations show that a significant
fraction of the baryons outside virialized regions have temperatures
in the range $10^5-10^7\; K$,  at least at the present time,  and that
the corresponding entropy is comparable, if not somewhat greater, 
than the minimum entropy level required to explain the various X-ray 
correlations exhibited by groups and clusters.  
Cen \etal~(1995) and Cen  \& Ostriker (1999) have suggested that this 
warm-hot diffuse medium is the result of the intergalactic medium
being shocked during  the formation of transient large-scale
features such as sheets and filaments.  A recent study by Dav\'e et 
al (2001) of the distribution of the warm intergalactic medium supports 
this hypothesis.  The implication of these results is that the 
entropy floor is a natural outcome of the currently favoured 
theoretical models. 

Finally, the UV, radio and X-ray observations are increasingly 
showing that the intracluster medium is not a simple single-phase 
thermal medium as has been generally assumed but rather,
is a  rich, complex phenomenon, much like the interstellar medium. 
Specifically, there is growing evidence that the intracluster medium
is composed of an X- ray luminous thermal component as well as a 
relativistic plasma --- comprised of relativistic electrons, tangled
magnetic fields and possibly even relativistic protons --- ejected
from quasars and active galactic nucleii (see, for example, 
Ensslin \etal 1997).
The presence of protons dramatically affects the evolution of the
plasma and the extent to which it can affect the thermal component.
Estimates of
the energetics suggest that the total energy of such a plasma will
be comparable to that of the thermal component (Ensslin \& Kaiser 2000).
Moreover, the presence of the protons ensures that the plasma will 
retain the bulk of its energy over cosmological
timescales even though the electrons will radiate away their energy 
via synchotron/inverse-compton processes on very short timescales.
The bubbles of relativistic plasma are expected to inflate until they 
reach pressure equilibrium with the ambient medium, in the process 
displacing the ambient thermal gas and creating low density cavities.  
There is growing evidence for the presence of such cavities in the 
X-ray observations of clusters \cite{Bohringer+95,Clarke+97,McNamara+2000}. 
The detection of $\sim 5--10\;\mu$G magnetic fields via Faraday rotation 
measurements in over 16 clusters (Feretti \etal 1995; 1999; Clarke, Kronberg,
Bohringer 2001) lends further credence to the scenario.  
The presence of a relativisitic fluid with substantial pressure in the 
protogroup/protocluster environment will modify 
the accretion flow onto the haloes,  especially those of lower mass.
It will also affect the equilibrium distribution of gas in the haloes, 
resulting in lower gas densities in the central regions and hence, an 
``entropy floor''.

In this paper, we explore the consequences of entropy injection by
one of the processes described above into the volume encompassing the
intergalactic medium {\it destined} to form the ICM of groups and
clusters.  Drawing upon insights derived from
detailed high resolution numerical simulations (\eg Lewis \etal 
2000) as well as 
building on prior work by Balogh, Babul \& Patton (1999), we have 
developed a simple, intuitive, physically motivated, {\em analytic} model that 
successfully captures most of the important physics associated with the 
accretion of 
high entropy gas onto the entire range of relevant halos, from low-mass groups 
to massive clusters.

A brief overview of the model is as follows:
Unaffected by the energy injection, 
the dominant dark matter component will, in due course, collapse and
virilize to form bound halos  (\S~\ref{model-dark}).
We assume that the dark matter in the
halos will settle into a cuspy distribution as suggested by  recent
high-resolution numerical simulations. On the other hand, the collapse
of the baryonic component is defined by the competition
between gravity and  the pressure forces engendered by the pre-heating
(\S~\ref{model-gas}).  If the  maximum infall velocity  purely due to 
the gravity of the halos ($\sim \sqrt{GM_h/R_h}$) is  subsonic, the 
flow will be strongly modified by pressure and the gas will not 
experience accretion shocks.  We assert that the baryons will accumulate
onto such halos isentropically at the rate given by the adiabatic
Bondi accretion rate.  This element of our model was first described in
Balogh, Babul, Patton (1999).   The treatment presented there, however, 
was restricted to low mass halos because the ``isentropic accretion'' 
assumption is only valid for such systems.   If the gravity 
of the halos is strong enough to drive the flow into the transonic or 
the supersonic regime,  the gas will experience accretion shocks and 
the concomitant increase in entropy.  In this paper, we extend our
earlier model by taking our cue from recent high-resolution hydrodynamic 
simulations of clusters and modeling the entropy profile of the gas 
as $S(r)=S_O + \alpha\ln(r/r_c)$ when the ``isentropic accretion'' 
assumption breaks down.   Under all conditions,  
the distribution of hot diffuse gas inside the halos is governed 
by the requirement that it be in thermal pressure-supported hydrostatic 
equilibrium within the halo's gravitational potential well.

The strength of our model lies 
in the fact it is both physically illuminating and allows us to compute and track the 
time-evolution of the X-ray properties of groups and 
clusters with relative ease.  Our model neglects the complicated process of gas cooling
but we note that, if the gas is preheated, gas cooling becomes less important 
because the
gas density in low mass systems is reduced.  

The present paper is organized as follows: 
In \S~\ref{sec-model}, we briefly review the Balogh, Babul, Patton (1999) model 
and describe the extension to cluster
scales, where accretion shocks become important.  
For specificity, we shall develop the model within the context of a flat $\Lambda$-CDM
cosmological model with $\Omega_{\rm m}=0.3$, $h=0.75$
and unless otherwise specified, a big bang nucleosynthesis value $\Omega_b=0.019 h^{-2}$ 
\cite{BBN,BBN2}.  
In \S~\ref{sec-results}, we first present the gas distributions in our models, and the
baryon fractions as a function of radius (\S\ref{sec-gasdist}).  We then explore
various scaling relations between gas temperature,
lumionsity, and cluster mass or velocity dispersion, and make extensive comparisons with
available data (\S\ref{sec-lt}--\S\ref{sec-ls}).  We also present temperature/luminosity functions for objects spanning
the entire mass range from groups to clusters, out to $z=1$, which demonstrate a very
favourable comparison with the available data (\S\ref{sec-LTF}).  
In \S~\ref{sec-discuss},  we discuss the results in light
of recent observations and critically assess the observational evidence favouring  
the ``pre-heated model''.  Our conclusions are summarized in \S~\ref{sec-conc}.

\section{The Theoretical Framework}\label{sec-model}

Our model for the formation of groups and clusters can be separated into two elements, which 
for simplicity's sake we treat separately.  The first element is the assemblage of the 
gravitationally bound structures (haloes) of the appropriate mass.  This process is essentially
 driven by the collapse and virialization of the dark matter.  The second element of the model
involves the accumulation and the subsequent redistribution, within the halo, of the diffuse 
baryons.  

\subsection{The Dark Matter Haloes}\label{model-dark}

Since dark matter is the gravitationally dominant component, we will assume that the first of the
two processes mentioned above proceeds independently of the second.  Specifically, we 
assume that the formation of the dark haloes as well as their internal structure and dynamics 
following virialization is everywhere dominated by the dark component.  

Let us first consider a population of haloes of mass $M_h$ observed at redshift 
$z_{\rm obs}$.  These haloes will have formed over a range of redshifts.  This distribution 
of the redshift of formation of a population of haloes of a given mass can be derived
using the analytic distribution function of Lacey, Cole (1993; 1994), assuming a
spectrum of initial density fluctuations given by
the Cold Dark Matter power spectrum \cite{BBKS}.  To do so, however,
we need to specify what we mean by ``formation''.  Following Balogh, Babul, Patton (1999), 
we define the epoch of 
formation of a given halo as the redshift, $z_{\rm form}$, when 75\% of the mass of the 
final halo mass at $z_{\rm obs}$ has been assembled and virialized in a halo.  We note
that throughout this paper, we will refer to the radius of {\em this} virialized region 
as the virial radius, ($R_{\rm vir}$), of the halo with the final (observed) mass 
$M_h$ that forms at redshift $z_{\rm form}$.  

The remaining 25\% of the halo's final mass accumulates between redshifts $z_{\rm form} 
< z < z_{\rm obs}$.  We  assume that this additional mass accretes gently --- so that 
internal structure of the halo is not disturbed --- and that it largely accumulates in 
the outer regions of the halo.  We will refer to the actual radius that encompasses mass 
$M_h$ at the epoch of observation ($z_{\rm obs}$) as $R_h$.

The above definition of halo formation is motivated by the results of numerical 
simulations \cite{NFW,NFW2} that suggest that the depth of the potential well, as 
traced by the circular velocity $V_c$ at the virial radius, remains relatively 
unchanged after $\sim$75\% of the cluster mass is in place;  the remaining 25\% of 
the mass is accreted typically in minor mergers that do not significantly disrupt 
the mass distribution already in place.   

Turning our attention to individual haloes, we assume that at $\zform$ the 
distribution of the virialized mass is given by:
\begin{equation}
\rho_{DM}(r)={\rho_{DM,\circ} \left({r \over r_s}\right)^{-n} \left(1+{r \over r_s}\right)^{n-3}},
\end{equation}
where $n=1.4$--$1.5$, $r_s$ is the scale radius and  $\rho_{DM,\circ}$ is the 
normalization of the profile.  Recent ultra-high resolution  numerical simulations 
\cite{Moore+98,Klypin,L+00}
show that the radial distribution of dark matter in the haloes is best 
described by such a profile.  

The normalization, $\rho_{DM,\circ}$, the size of the virialized region, 
$R_{\rm vir}$, 
and therefore, the shape and the depth of the gravitational potential well 
of the 
haloes will vary according to the epoch of halo formation; haloes of a given mass 
that form at an earlier epoch are denser, more compact and therefore, have a 
deeper potential.  
The normalization and the size of the virialized region at $\zform$ ($R_{\rm vir}$) 
are  specified by the requirements that (1) the mass contained {\em within} 
the virial radius is $M_{\rm form}=0.75 M_h(z_{\rm obs})$ and (2) the mean 
density interior to $R_{\rm vir}$ satisfies the virialization criterion:
\begin{align}
\bar{\rho}_{DM}(R_{\rm vir}, z_{\rm form}) & 
             =\rho_{DM,\circ}(z_{\rm form})\left({3\over 3-n}\right) \left({r_s\over R_{\rm vir}}\right)^n \;\;\ \ \ \ \ \ \ \ \ \ \ \   \notag\\ 
&\ \ \ \ \times\  _2F_1(3-n,3-n,4-n,-R_{\rm vir}/r_s)\notag \\
& =F(z_{\rm form})^2 \Delta_c(0)\rho_{\rm crit}(0),
\end{align}
where $_2F_1$ is the hypergeometric function and [c.f. Balogh, Babul, Patton (1999) for 
details] 
\begin{align}
F(z)^2 =& \left[1-\Omega_m+(1+z)^3\Omega_m\right]\Delta_c(z)/\Delta_c(0),\notag\\
\Delta_c(z) =& 49+96\Omega_m(z)+{200\Omega_m(z) \over 1+5\Omega_m(z)},\notag\\
\Omega_m(z) =& {\Omega_m(1+z)^3 \over 1-\Omega_m+\Omega_m(1+z)^3}.\notag
\end{align}

As for the scale radius, ultra-high resolution numerical simulations
of cluster-scale haloes indicate that $r_s/R_{\rm vir}\approx 0.20$--0.25
\cite{L+00}.  We, therefore, assume $r_s=$0.25$R_{\rm vir}$.  
There are suggestions \cite{NFW2,B+00} that this ratio may vary 
weakly with mass, with the ratio $r_s/R_{\rm vir}$ being somewhat smaller
for lower mass systems.  We have --- for present purposes --- chosen to 
ignore this complication.  We do not expect the distribution of preheated 
gas in our model to be sensitive to the precise value of $r_s$ because the 
gas temperatures is much hotter than the ``temperature'' corresponding to the
local gravitational potential.

\subsection{The Hot Diffuse Gas in The Haloes}\label{model-gas}

The fundamental assertion underlying our model is that one of the processes described 
in \S\ref{sec-intro} injects energy into the volume encompassing matter {\em destined} to collapse
to form groups and clusters of galaxies, raising the entropy of the diffuse gas
in the volume such that $kTn_e^{-2/3}$ is in the range $100$--$500$ keV cm$^{2}$.
For the present purposes,  we shall assume that the added energy increases
the thermal energy of the gas (as opposed to introducing a relativistic component).
For simplicity, we also assume that there exists a single, universal, initial  value 
of $kTn_e^{-2/3}$ and attempt to ascertain its value from the observations.  

As described in Balogh, Babul, Patton (1999), the response of the heated diffuse gas 
to the gravitational 
collapse and virialization of the group/cluster dark matter halo is determined 
by the competition between the opposing forces of pressure and gravity.   
In the case of sufficiently small haloes, the pressure forces can slow down 
the gas accretion flow to the point where the flow is subsonic everywhere and 
these haloes will not be able to 
accrete their full complement of baryons, $M_{\rm b}=(\Omega_{\rm b}/ \Omega_{\rm m})\; 
M_h$, by $z_{\rm obs}$.  For such systems, we follow the prescription 
outlined in Balogh, Babul, Patton (1999).  We assert that the gas distribution will accrete isentropically 
onto the haloes, and continue to evolve adiabatically.  The equation of state of a gas distribution 
evolving thusly is $P=K_\circ \rho_{\rm gas}^{5/3}$, where the constant $K_\circ= 
kT /(\mu m_H\rho_{\rm gas}^{2/3})$ (where $\mu=0.59$ for fully ionized H 
and He plasma with Y=0.25) is a measure of the specific entropy of the gas
and the accretion rate of the gas onto 
the halo is specified by the adiabatic Bondi accretion rate \cite{Bondi}:

\begin{align}\label{eqn-ABR}
\dot{M}_{\rm Bondi}&= 4\pi\lambda G^2M_h^2(\gamma K_\circ)^{-3/2}
\rho_g^{{3 \over 2}(5/3-\gamma)}, \\
       &\approx1.86\pi\lambda G^2M_h^2K_\circ^{-3/2}, \text{\ \ \  $\gamma=5/3$}\notag
\end{align}
where $\lambda=0.25$ is the dimensionless accretion rate.  For completeness, we note that 
we shall often quote $K_\circ$ in units of $K_{34}\equiv 10^{34}\ergs\g^{-5/3}\cm^2$, which,
for a fully ionized H and He plasma with Y=0.25, 
is also equivalent to $kTn_e^{-2/3} =948.9\kev\cm^2$.

As discussed in Balogh, Babul, Patton (1999), the gas content of the halo at $z_{\rm obs}$ can be estimated as
$M_{\rm gas}\approx \dot{M}_{\rm Bondi} t_H(z_{\rm obs})$, where $t_H$ is the Hubble time.
Consequently, the halo gas fraction will scale as $M_{\rm gas}/M_h \propto M_h$. 
There is, however,  a threshold mass above which the above estimate for $M_{\rm gas}$ implies
$M_{\rm gas}/M_h > \Omega_{\rm b}/\Omega_{\rm m}$, \ie a halo gas fraction that is
larger than the universal baryon fraction.  For such haloes, we cap the gas fraction at the 
universal value, recognizing that gas cannot fall in at a rate faster than that dictated 
by gravity.  We also note that in this discussion of the halo gas fraction, we have 
implicitly assumed that the fraction of baryons locked up in stars is, to first order, 
negligible (\eg Fukugita, Hogan, Peebles 1998, Balogh \etal 2001) and therefore, $M_{\rm gas}\approx M_{\rm b}$.

In Balogh, Babul, Patton (1999), we determined the distribution of gas in the haloes by requiring 
that the isentropically accreted gas, with equation of state 
$P=K_\circ\rho_{\rm gas}^{5/3}$, is in thermal pressure-supported hydrostatic
equilibrium within the halo's gravitational potential well.  This, combined
with the total amount of gas in the haloes, completely specifies the gas
density distribution in the haloes:

\begin{equation}\label{isentropic}
\begin{split}
{\rho_{\rm gas}(r) \over\rho_{\rm gas}(R_h)} &=
\left[1 + 
C
\int_{r}^{R_h}  \left({r'\over R_h}\right)^{-2} 
M_h(r') \; dr'\right]^{3/2},\\
&\ \ \ \ \ \ C={2G \rho_{\rm gas}(R_h)^{-2/3}  \over 5 K_\circ R_h^2 }
\end{split}
\end{equation}

However, the above treatment, like the assertion that the gas will accrete onto 
the haloes isentropically breaks down for sufficiently high mass 
haloes.  According to Equation~\ref{isentropic}, the density (and hence, 
the temperature) of the intracluster gas at the  halo radius decreases with 
increasing halo mass, and tends towards zero.  This result is clearly 
unphysical and indicative of the growing importance of accretion shocks.  

Physically, the accretion flow is expected to  behave as follows:
during accretion onto a low mass halo, the gas velocity never 
exceeds the local sound speed and the flow can be treated as isentropic.
However, as the mass of the halo is increased,  the gas pressure forces 
become progressively less important in comparision with the gravity and the maximum 
infall velocity of the accreting gas will rise from subsonic through trans-sonic to 
supersonic.  In the latter case, the gas  that falls onto the halo  
will experience accretion shocks that will vary in strength from weak shocks, associated
with mildly supersonic flows, to very strong shocks in the case of 
highly supersonic flows.  Since the passage through shocks is marked by an increase 
in entropy, the flow can no longer be considered isentropic.

Taking our cue from the above physical description and the 
the results of recent high resolution hydrodynamic simulations 
of cluster formation  \cite{L+00}, we model the distribution of gas in 
haloes where the gas has traversed through an accretion shock as follows: 
We demand that, regardless of how gas accretion occurs, the
temperature of the intracluster at the halo radius at $z_{\rm obs}$ 
be greater than, or equal to, $\frac1/2\; T_h$, where $kT_h\equiv
\frac1/2\; \mu m_H V_c^2(R_h)$ is the temperature corresponding to the
circular velocity at $R_h$.  

The above constraint establishes a critical halo mass: $M_{\rm isen}$.  For
$M_h < M_{\rm isen}$, the isentropically accreted gas in hydrostatic equilibrium within
the halo (with a density profile given by Equation~\ref{isentropic})
is, at $R=R_h$,  hotter than $\frac1/2\; T_h$.  We can, 
therefore, safely assume that the gas accreted isentropically as described in 
Balogh, Babul,  Patton (1999) and summarized above.  

For more massive haloes, the temperature of the isentropic distribution of gas
at $R_h$ is less than $\frac1/2 T_h$ and the gas
must therefore be shock heated in order to meet our temperature constraint.
To follow the shock history accurately requires a detailed treatment
of the halo merger history, which we do not consider.  However, at some early time, the 
most massive cluster progenitor will have had a mass
less than $M_{\rm isen}$, and will have accreted its gas isentropically.
We will assume, then, that a cluster is able to adiabatically accrete a total amount of
gas given by 
$(\Omega_{\rm b}/\Omega_{\rm m})\; M_{\rm isen}$,
and that this gas settles to
the bottom of the potential well, forming an isentropic core with a 
radius $r_c$.   Gas that accretes after the haloes have grown 
more massive than $M_{\rm isen}$ will be shocked (and shock-heated), 
with increasing strength as the halo grows more massive.

Let us consider a gas shell that accretes after the halo mass exceeds
$M_{\rm isen}$ but before it has grown to its final (observed) mass 
$M_h(z_{\rm obs})$.  As the gas shell is shocked, its entropy 
increases.  Once through the shock, numerical simulation results 
\cite{L+00} suggest that the gas shell  will evolve adibatically 
within the halo potential, sinking deeper into the potential, compressing 
and heating up as additional gas shells are accreted.  The gas in the 
shell will obey the relationship $P=K \rho_{\rm gas}^{5/3}$, where the 
value of $K$ is set by the post-shock value of the gas entropy in the shell.

Generalizing the above, we assume that even outside the isentropic core,
the gas equation of state is $P=K(r) \rho_{\rm gas}^{5/3}$, where 
$K(r)=K_\circ$ for $r \leq r_c$.  Again taking our cue from
the Lewis \etal (2000) simulation results, we model the entropy profile 
of the gas as
\begin{equation}
S(r)\equiv \ln \left({T / \rho_{\rm gas}^{2/3}}\right)=
S_\circ+\alpha \ln(r/r_c),
\end{equation}
where $\alpha=0$ for $r< r_c$, and therefore,
\begin{equation}
\ln K(r)=\ln K_\circ+\alpha \ln(r/r_c).
\end{equation}
Neglecting the self-gravity of the gas and 
requiring that the gas is in thermal pressure-supported hydrostatic equilibrium 
within the halo potential yields:
\begin{equation}
\begin{split}
\rho_{\rm gas}(r)&=\rho_{\rm gas}(r_c) \left({r\over r_c}\right)^{-3\alpha/5}\ \ \ \ \ \ \ \ \ \ \ \ \ \ \ \\
&\;\;\times\left[1 - 
C^\prime
\int_{r_c}^r  \left({r'\over r_c}\right)^{-(2+3\alpha/5)}
M_h(r') \; dr'\right]^{3/2}\!\!\!\!\!\!\!\! ,\ \ \ \\
&\ \ \ \ \ \ \ \ \ \ \ \ C^\prime={2G \rho_{\rm gas}(r_c)^{-2/3}  \over 5 K_\circ r_c^2 }
\end{split}
\end{equation}
where, as noted above, $\alpha=0$ for $r< r_c$ and $M_h(r)$ is the 
halo dark matter mass within radius $r$.  The gas temperature is given by:
\begin{equation}
kT(r)=\mu m_H K_\circ \left({r \over r_c} \right)^\alpha \rho_{\rm gas}(r)^{2/3}.
\end{equation}

Specifying the three parameters: $r_c$, $\rho_{\rm gas}(r_c)$ and $\alpha$
completely determines the model.  The normalization,  $\rho_{\rm gas}(r_c)$,
is determined by requiring that the total gas mass within $R_h$ is equal to 
\hbox{$(\Omega_{\rm b}/\Omega_{\rm m})\;M_h$}.  The core radius is set by requiring 
$M_{\rm gas}(r_c)=(\Omega_{\rm b}/\Omega_{\rm m})\; M_{\rm isen}$.
Finally, the value of $\alpha$ is constrained by requiring that $T(R_h)=
\frac1/2 T_h$. 

\begin{figure}
\begin{center}
\leavevmode \epsfysize=8cm \epsfbox{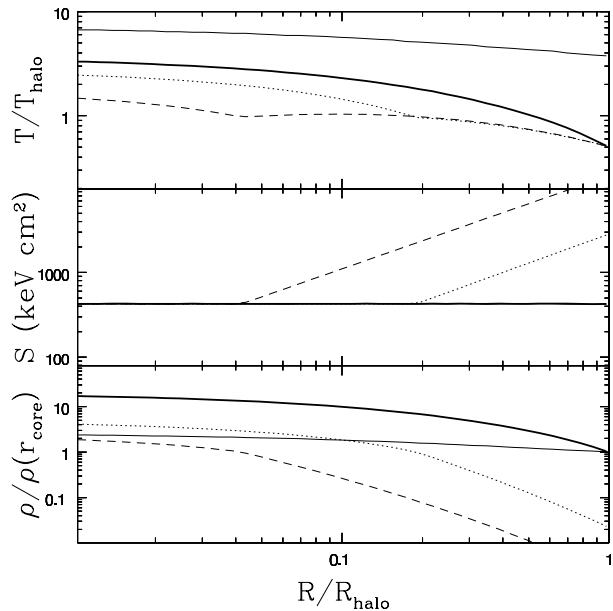}
\end{center}
\caption{The three-dimensional radial profiles of the gas
temperature, entropy and density for representative haloes at $z=0$, assuming that the 
entropy constant
of the preheated gas is $K_\circ=0.45 K_{\rm 34}$ or $kTn_e^{-2/3}\approx 427\kev\cm^2$.
The {\it thick, solid} curves show the profiles for a halo with critical mass $M_h
=M_{\rm isen}\equiv 8.4\times 10^{13}\msun$, 
\ie a halo where the gas is isentropic and its temperature at $R_h$ is equal to 
$\frac1/2 T_h $.  The {\it thin, solid} lines show the profiles for a halo with 
subcritical mass ($M_h<M_{\rm isen}$).  The {\it dotted} and the {\it dashed} lines 
correspond to the 
profiles in super-critical haloes of mass $M_h=4.75M_{\rm isen}=4\times 10^{14}\msun$ 
and $M_h=47.5M_{\rm isen}=4\times 10^{15}\msun$, respectively.  
}\label{profiles1}
\end{figure}

\begin{figure}
\begin{center}
\leavevmode \epsfysize=8cm \epsfbox{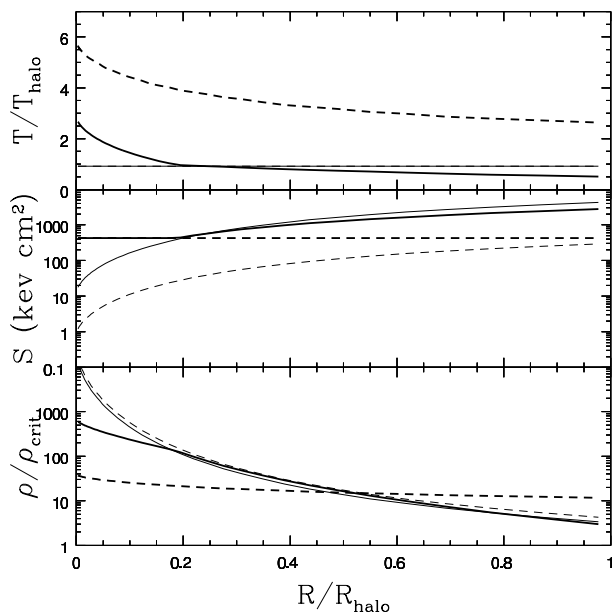}
\end{center}
\caption{A different perpective on the three-dimensional radial profiles of the gas
temperature, entropy and density for a group with $M_h=10^{13}\msun$ (subcritical, {\it dashed} curves)
and a cluster with $M_h=4\times 10^{14}\msun$ (supercritical, {\it solid} curves).  
The {\it thick}
lines correspond to the preheated model with $K_\circ=0.45 K_{\rm 34}$ or 
$kTn_e^{-2/3}\approx 427\kev\cm^2$ and the {\it thin} lines show the standard, isothermal
model with $T_{\rm gas}=T_{\rm vir}(z_{\rm form})$.  The juxtaposition of the 
preheated and the isothermal profiles illustrates the impact of the entropy floor
on the temperature and density profiles.}\label{fig-profiles2}
\end{figure}

\subsection{An isothermal model}
To compare and contrast the properties of the model described above, we also
construct a model in which the gas is assumed to be isothermal.  We use
the same halo potential, and again require the gas to be in pressure-supported
hydrostatic equilibrium, but at a temperature $T=T_{\rm vir}(z_{\rm form})\approx 
T_h$.  

\section{Results}\label{sec-results}

\subsection{Properties of the Gas Distribution in Model Groups and Clusters}\label{sec-gasdist}

In Figures \ref{profiles1} and \ref{fig-profiles2}, we plot the three-dimensional 
radial profiles of the gas temperature, entropy and density for a set of 
representative haloes at $z=0$.  In computing these, we have adopted a value of 
$K_\circ=0.45 K_{\rm 34}$ or equivalently, $kTn_e^{-2/3}\approx 427\kev\cm^2$ 
(for a fully  ionized H and He plasma with Y=0.25).  
This particular choice for the value of the entropy is the consequence of 
our requiring the model L--T relationship to match both the shape and the 
amplitude of the observed trend across the entire range of systems, from poor 
groups to rich clusters.  We discuss this more fully in \S~\ref{sec-lt}, where
we also compare our value to those adopted in other theoretical studies as
well as to those derived from X-ray observations.  We note that for our
adopted entropy level, the corresponding value 
of the critical halo mass is  $M_{\rm isen}=8.4 \times 10^{13} \msun$.  

The temperature, entropy and density profiles for a subcritical halo with mass 
$M_h = 3\times 10^{12}\msun$ are plotted as the thin, solid curves in 
Figure \ref{profiles1}.  The gas
has accreted onto the halo isentropically and therefore, the entropy profile is 
constant across the halo at a value of $kTn_e^{-2/3}\approx 427\kev\cm^2$.  
In addition, both the temperature and the density profiles are nearly flat,
within a factor of two.
We also note that the 
gas temperature at $R=R_h$ is nearly a factor of $\sim 4$ larger than 
$T_h$.  As seen in Figure \ref{fig-profiles2}, the impact of the entropy 
floor is to push the temperature of the gas that accumulates in subcritical haloes
well above the $T_{\rm vir}$ for the halo and, as a result, the gas is more diffuse
and extended compared to that of the isothermal model.

As the halo mass is increased, the profiles tend to steepen.  The profiles for a halo 
with critical mass $M_h=M_{\rm isen}=8.4\times 10^{13}\msun$ are shown as thick, 
solid curves in Figure \ref{profiles1}.  This is the most massive cluster onto which 
the gas is able to accrete 
isentropically, hence the flat entropy profile.  The temperature of the gas at 
$R=R_h$ is equal to $\frac1/2 T_h$.  This is also true (by definition) 
of the temperature of the gas at $R=R_h$ in all haloes more massive than 
$M_{\rm isen}$.  In Figure \ref{profiles1}, the profiles of two super-critical haloes 
of mass $4\times 10^{14}\msun$ and $4\times 10^{15}\msun$ are shown as dotted and 
dashed curves, respectively.  The gas
in these two supercritical haloes, however, is not purely isentropic.  Gas that accretes
onto haloes that eventually grow to be supercritical while their mass is subcritical,
will accrete isentropically and form the isentropic core.  Gas that accretes onto these 
haloes after their mass has grown larger than $M_{\rm isen}$ will experience accretion 
shocks and an associated increase in entropy.   An examination of the entropy profiles 
shows this core/envelope structure of the gas distribution in supercritical haloes.
If, in these massive haloes, we characterize the size of the isentropic core by its
radius $r_c$, then for a critical mass halo (short dashed curves), the core radius 
is equal to the radius of the halo and for supercritical haloes, $r_c < R_h$.
We point out that the ``kink'' in the density and temperature profiles for super-critical haloes
is a consequence of our treating the entropy profile as piecewise continuous.  In reality, 
we expect this transition to be smooth.

For supercritical haloes, the impact of the entropy floor diminishes with increasing
halo mass.  This is best illustrated by a comparison of the preheated and isothermal
profiles for a supercritical halo ($M_h=4\times 10^{14}\msun$); we show this
comparison in Figure \ref{fig-profiles2}.  Beyond
the central isentropic core, the temperature and density profiles of the two models
are very similar.  This is particularly true for the density profiles.  The gentle
decline in the temperature profile of the preheated model is, in fact, not due to 
preheating but rather due to compression of gas shells.  This type of compressional
heating is expected even if the gas had not been preheated.  Recent
cluster simulations also show a gently declining temperature profile towards the cluster
periphery;  in fact, outside the isentropic core, the temperature 
profile of the preheated gas model agrees very well with the results of the
Lewis \etal~(2000) simulation of a Virgo-mass cluster, which did not include
any preheating.  
As an aside, we note that this latter result suggests that, strictly speaking, a
purely isothermal model is not the most realistic standard with which to compare the
pre-heated models, though it is still a fairly good approximation.

We note that one possible way of testing some of the key assumptions
underlying our model, such as the universality of the value of the entropy floor,
is to compare the radial density, temperature and entropy profiles of the
gas distribution in our model halos against those of actual groups and clusters.
Recent observations, based on ROSAT PSPC observations of low-temperature groups,
suggest that entropy gradients in these groups are not flat 
(Lloyd-Davies, Ponman \& Cannon 2000), in apparent conflict
with the model presented here.  However, the systematic
uncertainties in determining even the
projected temperature gradient from these data are sufficiently large that
we cannot claim a strong discrepancy.   Observational data from {\it Chandra}
and {\it XMM}, which have superior spatial and energy resolution, will provide
a stronger test.  Furthermore, it is essential that the theoretical models
be subjected to the same inherent observational biases due to low surface
brightness, limited resolution and deprojection analyses, in order to make
a fair comparison with the data (see \S~4).
We are currently in the process of
carrying out analyses of this kind (Poole et al, in preparation)

\begin{figure}
\begin{center}
\leavevmode \epsfysize=8cm \epsfbox{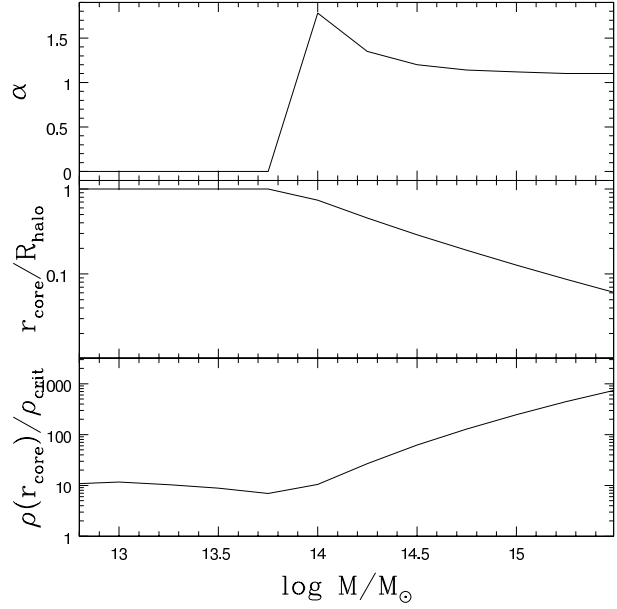}
\end{center}
\caption{A plot of the three model parameters, $\alpha$, $r_c$ and 
$\rho_{\rm gas}(r_c)$, in haloes of different mass. $\alpha$ determines the
entropy profile outside the core; inside the core, $\alpha=0$.  These parameters
correspond to an entropy floor of $K_\circ=0.45 K_{\rm 34}$ or 
$kTn_e^{-2/3}\approx 427\kev\cm^2$. 
}\label{model-param}
\end{figure}

Finally, for the convenience of readers who would like to reconstruct our model
profiles for their own use, we plot in Figure \ref{model-param}, the values of the 
three model parameters, $\alpha$, $r_c$ and $\rho_{\rm gas}(r_c)$,
in haloes of different mass.  We remind the reader that $\alpha$ determines the 
entropy profile for $r > r_c$; inside the core, the gas is isentropic 
and $\alpha=0$.  
Note that for clusters more massive than $M\approx 10^{14.5}\msun$,
the slope of the entropy profile outside the core
is nearly independent of cluster mass  (to within 10 per cent,
$\alpha\approx 1.1$) while $r_c$ steadily decreases.  This means that, 
outside the isentropic core, the entropy profiles of all clusters more 
massive than this limit are predicted to be nearly identical, once they 
are scaled to the value of the entropy at the virial radius.

\begin{figure}
\begin{center}
\leavevmode \epsfysize=8cm \epsfbox{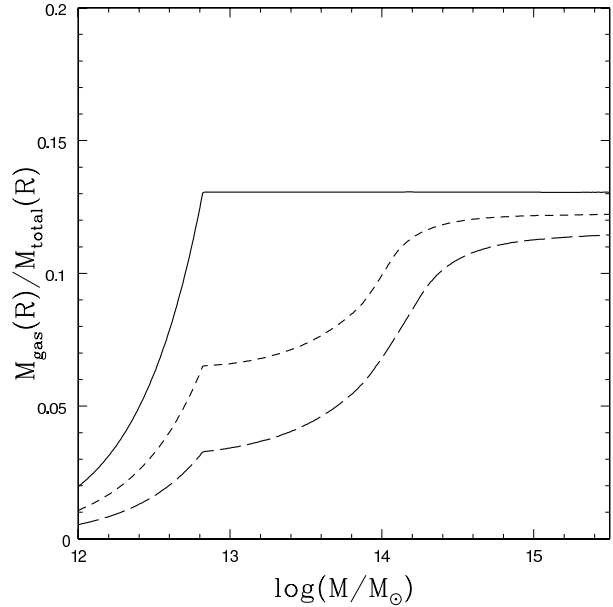}
\end{center}
\caption{The {\it solid, short-dashed} and {\it long-dashed }
curves give the mean gas fraction within 
three different radii ($R_h$, $R_{200}$ and $R_{500}$, respectively) in the 
haloes at $z_{\rm obs}=0$, as a function of the total halo mass.  
The $R_h$ curve shows the gas fraction within the entire halo.  
This fraction is constrained 
to be no greater than $M_{\rm gas}/M_h=0.112$, the universal value for 
the cosmology under consideration.  $R_{200}$ and $R_{500}$ are radii within 
which the mean dark matter mass density is 200 and 500 times the critical
cosmological density at $z_{\rm obs}=0$, respectively.
All of the curves are computed assuming 
that the entropy constant of the preheated gas is $K_\circ=0.45 K_{\rm 34}$ or 
$kTn_e^{-2/3}\approx 427\kev\cm^2$.  
}\label{gas-frac}
\end{figure}

In Figure \ref{gas-frac}, we plot the gas fraction $M_{\rm gas}/M_h$ within 
different regions characterized by $R_h$, $R_{200}$ and $R_{500}$ in the haloes 
at $z_{\rm obs}=0$, as a function of the total halo mass.  
These curves are computed using three dimensional gas and dark matter
density profiles; the gas profile corresponds to that for an entropy constant
$K_\circ=0.45 K_{\rm 34}$.
The short-dashed and long-dashed curves shows the gas fraction in regions within which the mean 
total mass density is 200 and 500 times the critical cosmological matter density 
at $z_{\rm obs}=0$, respectively.
Although the amplitudes of the two curves are different, their behaviour is similar.
For very low mass haloes ($M\lta 7\times 10^{12}\msun$), the gas fraction inside 
$R_{200}$ and $R_{500}$ is very small because the gas fraction in the entire halo 
is much less than the universal value of 0.0112 (c.f. the solid curve, which shows 
the gas fraction within the entire halo)
and thermal pressure prevents the gas that is in the halo from concentrating in the center.  
With increasing halo mass, the halo gas fraction rises as $M_{\rm gas}/M_h 
\propto M_h$.  The gas mass  rises faster than the total halo mass.  This
increase, coupled with compression of the gas, results in a rise in 
the gas fractions within $R_{200}$ and $R_{500}$.  
As per our model, once the halo gas fraction reaches the universal value, it cannot
increase any further, hence the flattening of the solid curve.  From this point 
on, the increase in $M_{\rm gas}$ tracks the increase in $M_h$.  The transition
also impacts upon the growth of the gas fraction within $R_{200}$ and $R_{500}$.  From
this point on, the increase in the gas fraction within these two radii is entirely
due to the compression and concentration of the gas.  The increase is only checked 
when the mass enclosed within the radius under consideration exceeds $M_{\rm isen}$.  For 
$R_{200}$, this happens at $M_h=1.3\times 10^{14}\msun$ and for $R_{500}$, 
this happens at a slighly larger mass of $M_h=1.6\times 10^{14} \msun$.

With an eye towards observations, the curves in Figure \ref{gas-frac} predict that
for systems with masses greater than $7\times 10^{12}\msun$, there should be 
no observed variation in the gas fraction within the halo as a whole.  However,
the gas fraction within $R_{200}$ or $R_{500}$ should increase steadily with
mass, approximately proportional to $M_h^{0.5}$ on mass scales below rich
clusters.
Recent analyses {\cite{DJF,Mohr,FT99,Bryan} 
suggest that the gas fraction in groups is lower that than in clusters, growing with
halo mass as $M_h^{0.4}$; however, this trend has yet to be firmly established.
As discussed by several authors (\eg Mulchaey 2000; Roussel \etal 2000; Balogh \etal 2001), 
while many rich clusters have their X-ray emissions
detected out to their peripheries, the detected X-ray emissions from groups is typically
limited to the central regions.  The lower gas fractions for groups, therefore, is more
likely indicative of the paucity of gas in the central regions.  

\begin{figure}
\begin{center}
\leavevmode \epsfysize=8cm \epsfbox{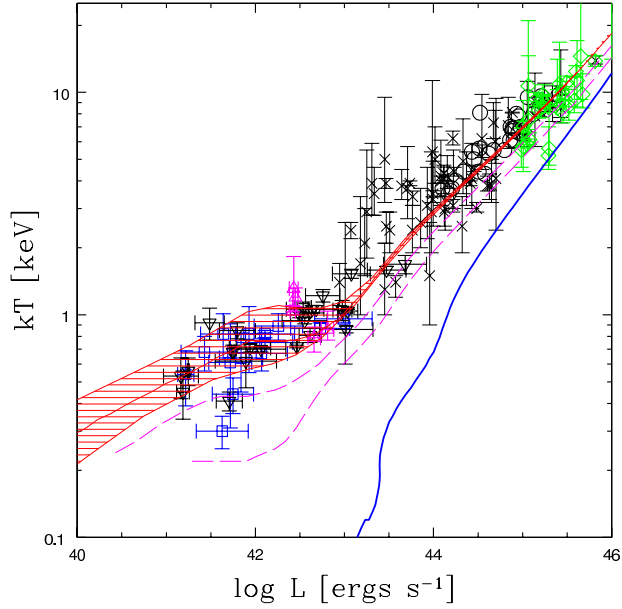}
\end{center}
\caption{The $z=0$ luminosity--temperature relation, with data from Markevitch (1998, {\it 
circles}), 
Allen, Fabian (1998, {\it diamonds}), David \etal (1993 ,{\it crosses}), Ponman 
\etal  (1996, {\it squares}), Mulchaey, Zabludoff (1998,  {\it triangles}), 
and Helsdon, Ponman (2000, {\it inverted triangles}).  The error bars are 
omitted from the Markevitch data, as they are smaller than the data symbols.
The {\it heavy, solid} curve is the L--T relationship for 
haloes with isothermal gas at temperature $T=T_{\rm vir}(z_{\rm form})\approx T_h$,
assuming the mean redshift of formation for haloes of a given mass.
The {\it hatched region} shows the model results for an entropy constant of $K_\circ=0.45 K_{\rm 34}$ 
or $kTn_e^{-2/3}\approx 427\kev\cm^2$.  The width of the region reflects the range of formation
 redshifts for clusters of a given mass.  Across the entire range, from poor groups to 
rich clusters, this ``preheated'' model results give an excellent match to observations.
Finally, the two {\it dashed} curves show the model
results for entropy constants $K_\circ=0.24 K_{\rm 34}$ and $K_\circ=0.12 K_{\rm 34}$
(\ie $kTn_e^{-2/3}\approx 227\kev\cm^2$ and $kTn_e^{-2/3}\approx 114\kev\cm^2$, respectively).
}\label{L-T}
\end{figure}

\subsection{The Luminosity-Temperature Relationship of Model Groups and Clusters}\label{sec-lt}

We use the latest version of the Raymond-Smith plasma code 
(Raymond, Cox, Smith 1976; Raymond, Smith 1977) to calculate 
the X-ray volume  emissivity of the hot gas in the haloes.
This code takes into account various  radiative processes that can
occur in  low density plasma such as permitted, forbidden, 
and semiforbidden line transitions, dielectronic recombination, 
bremsstrahlung,  radiative recombination, and two-photon continua. The
code can treat plasma of arbitrary metallicity, taking into account
the influence of  elements up to Fe and Ni, with temperature ranging from
 $10^4$ K to $10^8$ K.  This is important since the gas in the haloes of 
interest to us range from a few $\times 10^6$ K ($\sim 0.3\kev$) to 
$10^8$ K ($\sim 10\kev$).  The X-ray emission of a plasma with $kT < 4\kev$ 
is dominated by recombination radiation whereas emission of hotter
gas is largely due to bremsstrahlung.  For the purposes of calculating
the X-ray emissivity of the hot plasma, we assume a constant metallicity of 0.3 Z$_\odot$.

In Figure \ref{L-T}, we plot the $z_{\rm obs}=0$  luminosity--temperature (L--T) 
relation for groups and clusters.  The luminosity  is the bolometric X--ray lumionsity
of the halo and is computed by integrating the volume emissivity out to $R_h$,
while the temperature is a luminosity--weighted average temperature (c.f. 
Equation 24 of Balogh, Babul, Patton 1999).  We compare these results with data from 
from David \etal (1993), Ponman \etal (1996)\footnote{Only fully resolved observations 
are considered (\ie with a quality index of 1).}, Allen \& Fabian (1998), Markevitch (1998), 
Mulchaey, Zabludoff (1998)\footnote{We use the temperatures determined using the 
Raymond--Smith model with the metallicity fixed at half solar for all groups 
except NGC5846, for which this temperature is unconstrained.  In this case, we adopt 
the low metallicity determination.} and Helsdon \& Ponman (2000).  The Helsdon, Ponman 
luminosities have been corrected, using their published estimates for the correction 
factors,  for emission between the observed radius and the virial radius.   For the 
other group data, we have adopted an average correction (based on the Helsdon, Ponman 
analysis) of a factor of two in luminosity.  We discuss this bias further in 
\S\ref{sec-discuss}.

As a reference point, we show the results for the isothermal model as the heavy solid curve.
This illustrates the now well-known result that the isothermal L--T
curve is too steep to match the observations, even for clusters with temperatures
greater than a few keV.  In this temperature range, the relationship scales
as $L\propto T^2$.  Due to the dominance of recombination radiation at temperatures
of less than 4 keV, the relationship steepens even more, approaching $L\propto T$. 
Also note that, in contrast with Balogh, Babul, Patton (1999), the isothermal model 
is overluminous at
all temperatures.  This is a consequence of adopting a dark matter potential which
is very concentrated; in Balogh, Babul, Patton (1999), we assumed an isothermal 
potential with a flat core,
which greatly reduces the total luminosity.   We note further that haloes
with $kT\lesssim0.5$ keV will radiate all of their energy in a Hubble time in the
isothermal model, and, thus the approximation that cooling can be neglected
breaks down severely in such systems.

Figure \ref{L-T} also shows the L--T relationship for the preheated models 
with entropy constants $K_\circ=0.45 K_{\rm 34}$ (the hatched region),
$K_\circ=0.24 K_{\rm 34}$ and $K_\circ=0.12 K_{\rm 34}$ (the two dashed curves).
A quick comparison of the three curves shows that the normalisation depends on the 
initial entropy.  
The lower of the two dashed curves is constructed with an entropy constant of
$K_\circ=0.12 K_{\rm 34}$ or $kTn_e^{-2/3}\approx 112\kev\cm^2$, close to that 
suggested by Helsdon, Ponman (2000) and Lloyd-Davies, Ponman, Cannon (2000).  
Although this curve may perhaps be seen
to trace the lower envelope for the data, it is clear that it fails to match
the majority of the
observations across the entire range from poor groups to rich clusters.  

To account for both the amplitude and the slope of the L-T relationships
across the entire range from poor groups to rich clusters, 
we are required to consider models with entropy constants
higher than $kTn_e^{-2/3}\approx 100\kev\cm^2$.  The hatched region in 
Figure \ref{L-T} corresponds to a model with $K_\circ=0.45 K_{\rm 34}$ or 
$kTn_e^{-2/3}\approx 427\kev\cm^2$.  This ``preheated'' model result 
gives an excellent match to observations across the entire range, from poor 
groups to rich clusters, and it is on the basis of this match to the observed 
L--T data that we adopt the $K_\circ=0.45 K_{\rm 34}$ as our preferred model.

Our ``preferred'' value of the entropy floor is considerably higher than 
the value of $60$--$120\kev\cm^2$ (for $h=0.75$) derived  by Lloyd-Davies 
\etal (2000) from an analyses of {\it ROSAT} PSPC  observations of 
low-temperature groups.  As we have already noted  (and elaborate 
upon further in \S~4), comparing theoretical results with  those derived 
from X-ray observations of groups is not straightforward  due to the 
substantial systematic and statistical uncertainties in the observations 
resulting from the finite resolution of the X-ray telescopes and the 
intrinsic faintness of the groups.  For one thing, only the very central
regions of the systems are often detectable above the X-ray background 
(Mulchaey 2000; Roussel \etal 2000; Helsdon, Ponman 2000) and typically, 
even this central region tends to be defined by only a small number of 
X-ray photons.  As shown in Figure 11, even crudely modeling the limited 
X-ray extent reduces the required value of the entropy floor to $\approx 
330\kev\cm^2$.  However, even this latter value is still higher than the 
``observed value''.  We, however, do not expect to be able to reduce the 
entropy floor significantly below this latter value without,  for example, 
altering the dark matter profile of the halo and/or jettisoning the 
assumption that the value of the entropy floor is a universal constant. As 
illustrated in Figure 5, lowering the  universal value of the entropy floor, 
while keeping the halo dark matter density distribution consistent with that 
seen in high-resolution cold dark matter simulations, leads to a drop in 
the amplitude of the model L--T curve and results in a mismatch between 
the theoretical and observational L--T correlations on the cluster scales.  
On these scales, the observational biases that affect the group results 
are not an issue.  And neither are most of the key assumptions underlying 
our model, such as the ``Bondi approximation'', as these predominantly 
affect only the distribution of gas in the low mass halos.

Comparing our value of the entropy floor with those required in 
other theoretical
models, we note that Tozzi \& Norman (2001) require a ``high''
level of entropy injection, in the range of  $190$--$960\kev\cm^2$,
to match the observed L--T correlations at $\sim 0.5$--$2\kev$.
On the other hand, Cavaliere \etal~(1999) assume an entropy injection
that is comparable to the value measured by Lloyd-Davies \etal (2000),
and while their model L--T relationship is consistent with the 
observations on the group scale, as noted by Lloyd-Davies \etal
(2000), they fail to reproduce the slope of the L--T relationship
at high temperatures.  Finally, motivated by the results of 
Lloyd-Davies \etal (2000), da Silva \etal (2001) have recently 
carried out numerical simulations with a mean entropy floor of 
$\sim 80 \kev\cm^2$ but they find that the $z=0$ groups and
clusters in their simulation volume do {\it not} match the 
observed X-ray scaling relations.  The upshot of all this is
that there is considerable more work that needs to be done
in order to bridge theory and observations.  One possibility 
is that the entropy floor is a function of the halo mass (but 
see the discussion in the Appendix of Balogh, Babul \& Patton 1999)
and therefore, the value of the entropy floor on group 
scale is intrinsically lower than on the cluster scale.
Alternatively, the results may be indicating that the underlying
dark matter distribution is too cuspy.  Flattening the inner
profiles of the dark matter halos reduces the depth of the 
potential well and may result in the lowering of the 
entropy floor.  (In Balogh, Babul \& Patton 1999, we adopted
dark matter halos with flat inner cores and required an entropy
floor of $\sim 350\kev\cm^2$ as opposed to $\sim 430\kev\cm^2$
in the present study in order to match the observed
group L-T results without even taking into account any 
observational biases.)
Group and cluster data from {\it Chandra}
and {\it XMM}, with superior spatial and energy resolution,
are likely to go a long way towards resolving such issues.

The $K_\circ=0.45 K_{\rm 34}$
model L--T curve is plotted in Figure 5 embedded within a hatched region, the width 
of which is related to the redshift range $z_{\rm obs}=0 < z_{\rm form} < z_{1\sigma}$
within which 68.3\% of the haloes have formed.  This dispersion in the model L--T 
relation is also in excellent agreement with the observed scatter.  
The width of the hatched region is 
larger at low luminosities than at high luminosities, reflecting the broader distribution
of formation times for low mass haloes [c.f. Balogh, Babul, Patton (1999), especially 
Figure 2, for details].  
Interestingly, this  1-$\sigma$ band encompasses most of the group data shown in 
Figure \ref{L-T}.  As noted by Balogh, Babul, Patton (1999), this match raises 
an extremely interesting possibility that the observed dispersion in the L--T relation 
is primarily due to the distribution of halo formation times.
The model also predicts a very small dispersion at high luminosities and several studies 
\cite{F+94,AE98,M98} have demonstrated that scatter in the L--T relation can be reduced 
to a very small value (an r.m.s. dispersion of about 0.11 in log L at a given T) by 
excluding cooling flow regions from the observed X-ray data.  It is also worth
pointing out that because of the very narow width of the hatched region at high 
luminosities, the $K_\circ=0.45 K_{\rm 34}$ matches the observations in this 
regime.

At low temperatures and luminosities, the ``preheated'' model 
L--T relation scales as as $L\propto T^{4.7}$.  This scaling
is in excellent agreement with the results of Helsdon \& Ponman (2000)
based on their sample of 24 X-ray bright groups with temperatures
ranging from $0.5$--$1.7\kev$.  It is also consistent with the 
results of Mulchaey \& Zabludoff (1998).  The latter claim that 
their data is consistent with $L\propto T^3$ that describes the 
cluster data, which it is, but the sample consists of only nine
groups spanning a rather narrow range in temperature.

At $L\approx 10^{42}\ergsec$ ($T\approx 0.8\kev$), 
the L--T curve veers away from the low luminosity, low temperature asymptote.
This point corresponds to the transition between haloes with 
$M_{\rm gas}/M_h < \Omega_{\rm b}/ \Omega_{\rm m}$ and 
$M_{\rm gas}/M_h = \Omega_{\rm b}/ \Omega_{\rm m}$.
Prior to this transition point, the gas density and temperature gradients in the
haloes are fairly shallow.  
Once past the transition point, the 
central gas density (and temperature)  increases while that at the halo radius 
drops.  The profiles, though steepening, are still sufficiently flat that the 
contribution of the hotter gas in the center to the emission-weighted temperature
is offset by the cooler gas at larger radii, and the observed temperature 
remains roughly constant.  In due course, however, the steeping of the gas density 
profiles with increasing halo mass begins to affect the emission-weighted temperature.
As the gas profile steepens, an increasing fraction of the halo luminosity comes 
from the central regions where the gas is hotter and once  a significant
fraction of the luminosity originates in these central regions, the observed 
temperature begins to rise.  Luminosity, on the other hand, depends 
sensitively on gas density and therefore, continues to increase throughout.

A second transition occurs at $L\approx 10^{43}\ergsec$ ($T\approx 1\kev$).  This
transition separates systems onto which the gas accretes isentropically and systems in which
accretions shocks become increasingly important.  Shock heating imposes a lower 
limit to the gas temperature.  As the halo mass grows larger and the shocked gas
comes to dominate that intracluster medium, the L--T systems asymptotes to 
$L\propto T^{2.5}$, close to the observed $L\propto T^{2.7}$ 
relationship for clusters \cite{ES91,MFSV}.

\begin{figure}
\begin{center}
\leavevmode \epsfysize=8cm \epsfbox{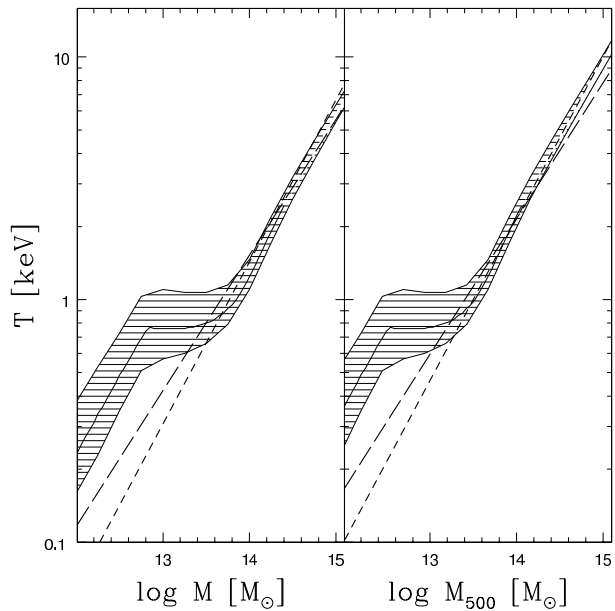}
\end{center}
\caption{
The $z=0$ mass-temperature relation for the ``preheated'' model with
entropy constant of $K_\circ=0.45 K_{\rm 34}$ or $kTn_e^{-2/3}\approx 427\kev\cm^2$.  
The mass in the {\it left panel} is the total halo mass while in the {\it right panel}, it 
corresponds to the mass within $R_{500}$.
The {\it hatched region} illustrates the dispersion expected due to the range for formation 
redshifts for groups and clusters of a given mass.  Over the entire range, from very low
to very high mass haloes, $M\propto T^{1.5-1.8}$, with the exact value of the power-law
index depending on which mass is being considered.  The {\it short dashed} curves show the
$M\propto T^{3/2}$ relationship expected from self-similar scaling arguments.  
The self-similar model has been renormalized to match the pre-heated results in 
the $T\approx 2-3\kev$ range.  For systems with $T < 0.7\kev$, the normalization of 
the pre-heated model is nearly a factor of $\sim 3.5$ higher than the power-law
extension of the high-temperature relationship.  Nevalainen \etal (2000) find that fitting
the observed mass--temperature data for systems with temperatures greater than 1 keV with 
a simple power-law yields $M\propto T^{1.79}$ ({\it long dashed} line).
}\label{M-T}
\end{figure}

\subsection{The Mass-Temperature Relationship}\label{sec-mt}

In Fig. 6, we plot the $z_{\rm obs}=0$  mass--temperature relation for 
groups and clusters for our preferred preheated model.  The temperature 
plotted is, as before, the emission-weighted temperature, and the mass 
in the left plot is the total halo mass while in the right plot, it 
corresponds to the mass within $R_{500}$.  A visual inspection of the 
figure shows that mass--temperature relationship is very sensitive to 
the thermal history of the gas.

For haloes with temperatures $T>1\kev$, the temperature rises monotonically 
with halo mass.  The $\sim 1\kev$ scale separates systems onto which the 
gas accretes isentropically and systems in which accretion shocks become 
increasingly important.  In the preheated model, the M--T relation for 
massive clusters is $M\propto T^{1.6-1.7}$, somewhat steeper than the 
self-similar result, if the mass being considered is the total mass, which 
is probably the best assumption for massive clusters (Mulchaey 2000).  The 
slope is indistinguishible from the self-similar result if one scales 
against $M_{500}$.  Ettori, Fabian (1999) as well as Nevalainen \etal (2000) 
have fit the mass-temperature data for systems with temperatures greater 
than 1 keV with a simple power-law and find $M\propto T^{1.7-1.9}$, in 
reasonable agreement with the model.

For low temperature groups ($T< 0.7\kev$), the preheated model predicts 
$M\propto T^{1.6}$, which is slightly steeper than the isothermal, self-similar 
$M\propto T^{3/2}$ relationship but similar to the M--T relation for massive 
pre-heated clusters.  Note, however, that the emission-weighted gas temperature 
of a group of a given  mass in the pre-heated model is nearly a factor of 3.5 
higher than that expected from the extension of the cluster M--T relationship 
to lower masses.  

As already discussed, in the preheated model the baryonic fraction of the low 
temperature groups is determined by the Bondi accretion rate and when this 
fraction exceeds the universal value, we argue that the baryon fraction of 
the halo will freeze at the universal value since gas cannot fall into the 
halo at a faster rate than that established by gravitationally induced 
accretion.  As in the case of the  L--T relation, this transition results in the 
model M--T curve breaking away from its low luminosity, low temperature asymptote 
at $T\approx 0.7\kev$.  For a range of masses, the emission-weighted temperature 
of the gas is nearly independent of the halo mass.  The reason for this behaviour
has already been discussed in the context of the L--T relationship (Section 3.2).

\subsection{The Luminosity-$\sigma$ Relationship}\label{sec-ls}

Since it is not possible to directly observe cluster masses, we consider
the line-of-sight velocity dispersion for the model haloes.  We
define this as $\sigma_{\rm los}=V_c/\sqrt{2}$, which is a good approximation for haloes
of our adopted potential \cite{LM00}.  
In Figure \ref{L-Sig}, we plot the $z_{\rm obs}=0$  $L$--$\sigma_{\rm los}$ relation 
for the groups and clusters.  We also show variety of sources
compiled from the literature:
Mulchaey, Zabludoff (1998), David \etal (1993), Ponman \etal (1996),
Markevitch \etal (1998),  and Xue, Wu (2000).

\begin{figure}
\begin{center}\leavevmode\epsfysize=8cm
\epsfbox{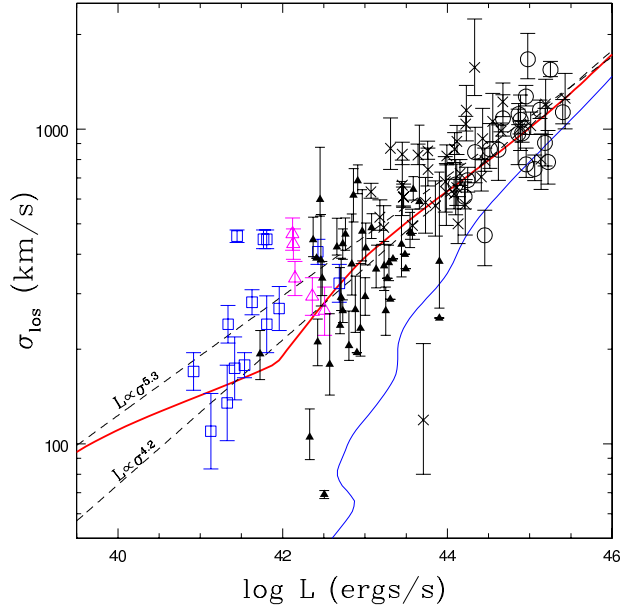}
\end{center}
\caption{The {\it thick solid} curve shows the
$z=0$ $L$--$\sigma_{\rm los}$ relation for the ``preheated'' model with
entropy constant of $K_\circ=0.45 K_{\rm 34}$ or $kTn_e^{-2/3}\approx 427\kev\cm^2$.
We also show $L-\sigma_{\rm los}$ relation ({\it thin solid} curve) for the isothermal, 
self-similar model.  The preheated model scales as  $L\propto\sigma_{\rm los}^{5}$
at high luminosities, and $L\propto\sigma_{\rm los}^{9.4}$ at low luminosities.
We also plot as {\it dashed lines} the power-law 
best-fit scaling found by Wu \etal  (1999, $L\propto\sigma_{\rm los}^{5.3}$) and
Mulchaey, Zabludoff (1998, $L\propto\sigma_{\rm los}^{4.2}$).
The data plotted are from Mulchaey, Zabludoff (1998, {\it open triangles}),
David \etal (1993, {\it crosses}), Ponman \etal (1996, {\it open squares}),
Markevitch \etal (1998, {\it circles}), and Xue, Wu (2000, {\it filled 
triangles}).
}\label{L-Sig}
\end{figure}

The thick solid curve shows the $z=0$ $L$--$\sigma_{\rm los}$ relation 
for the ``preheated'' model with entropy constant of $K_\circ=0.45 K_{\rm 34}$ or 
$kTn_e^{-2/3}\approx 427\kev\cm^2$.  At high luminosities, the curve asymptotes to 
$L\propto\sigma_{\rm los}^{5}$ and at low luminosities, it scales as 
$L\propto\sigma_{\rm los}^{9.4}$.  Given the considerable scatter in the data, 
the ``preheated'' model prediction is indistinguishable from the power-law scaling 
relations $L\propto\sigma_{\rm los}^{4.2}$, $L\propto\sigma_{\rm los}^{4.4}$,
$L\propto\sigma_{\rm los}^{4.5}$, $L\propto\sigma_{\rm los}^{4.9}$ and 
$L\propto\sigma_{\rm los}^{5.3}$ of Mulchaey, Zabludoff (1998), Mahdavi \& Geller 
(2001), Helsdon \& Ponman (2000), Ponman et al (1996) and Wu \etal (2000), respectively.  
In fact, analyses of the group and cluster data by Ponman et al (1996), 
Mulchaey, Zabludoff (1998), Helsdon, Ponman (2000) and Mahdavi, Geller (2001) all 
agree that as it stands, the $L$--$\sigma_{\rm los}$ relation for the groups is 
essentially  the same as that for the clusters.  As noted by Mulchaey (2000), 
``within the errors, the slopes derived by Mulchaey, Zabludoff(1998), Ponman et 
al (1996) and Helsdon, Ponman (2000) are indistinguishable.''

\subsection{The Group-Cluster  Temperature and Luminosity Function}\label{sec-LTF}

The mass function of dark matter haloes is now quite well established
by theory, and depends on cosmology and the shape of the power spectrum.
Although Press-Schechter formalism, which has been used by many authors
(including ourselves in Balogh, Babul, Patton 1999), provides a (surprisingly) good description
of the mass function, large numerical simulations (\eg Governato \etal 1999)
now provide us with a firm basis for the development of an improved
formalism.  We use the mass function of Jenkins \etal (2001),
which is a good, ``universal'' description, to within about 10\%.  We 
leave the normalisation, $\sigma_8$, as a free parameter.

The differential luminosity function is obtained from the mass function
and the derivative $dM/dL$, which we evaluate numerically.  An
analagous procedure provides us with the differential temperature function;
for comparison with observations, we integrate this function to give the cumulative temperature
function, the number of galaxies with temperatures greater than $T$.  

\subsubsection{The Temperature Function}
In Figure \ref{fig-Tfunc} we show the observed temperature function from Henry (2000),
after making the necessary volume correction to convert the results to our adopted
cosmological model.  The solid points are the data at $z=0$.  The open symbols represent
clusters at $z\approx 0.4$, and show clear evidence for negative evolution;
clusters of a given temperature are about three times less common at $z=0.4$ than they
are locally.

\begin{figure*}
\begin{center}
\leavevmode \epsfysize=8cm \epsfbox{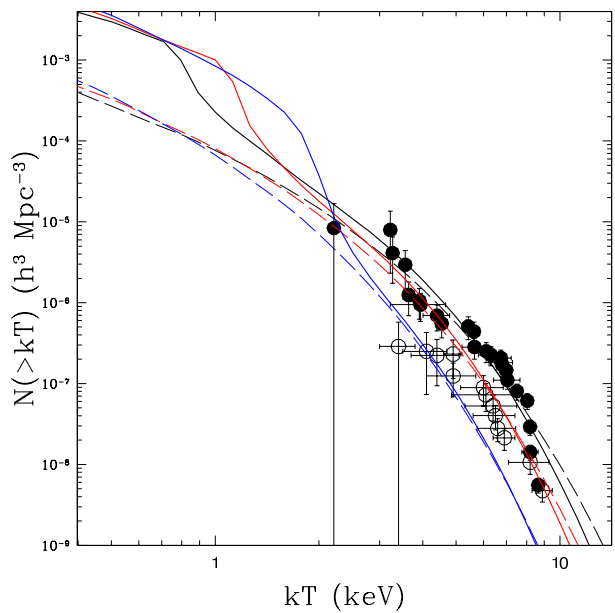}
\end{center}
\caption{The cumulative temperature function.  Data are from Henry (2000), converted
to our cosmological parameters.  The {\it solid points} are the $z=0$ data, and the
{\it open points} are clusters at $z\approx 0.4$.  The {\it dashed lines} are the isothermal
models at $z=0$, $z=0.4$, and $z=1$, such that $N(>T)$ decreases with redshift at
high temperatures.  The {\it solid lines} are the preheated models at the same redshifts;
the direction of evolution at the high temperature end is the same as for the isothermal
models.  We choose $\sigma_8=0.88$ and $\sigma_8=0.78$ for the isothermal and preheated
models, respectively, to provide a good match to the $z=0$ data.
\label{fig-Tfunc}}
\end{figure*}

The isothermal model results are plotted as dashed lines in Figure \ref{fig-Tfunc}.  We have chosen
$\sigma_8=0.88$ so that the model provides a reasonable match to the $z=0$ data.
Model results at $z=0.4$ and $z=1.0$ are also plotted, and they demonstrate negative
evolution for $kT>1$ keV.  The direction and amount of evolution to $z=0.4$ in the
data is well matched by the model.

The solid lines in Figure \ref{fig-Tfunc} represent the temperature function derived from our
preheated models.  In this case, we need to use a lower value of $\sigma_8=0.78$ to
match the $z=0$ data.  Thus, the normalisation of the power spectrum from the
temperature function is dependent on the thermodynamic history of the gas, to about
10\%.  At high temperatures, $kT>3$ keV, the direction and amount of evolution of these
models is quite similar to the isothermal models.  Though the shape of the temperature
function itself is steeper than that of the isothermal, it is not at the level which will
be easily measured observationally.  Therefore, apart from the normalisation,
it is difficult to distinguish between the models on cluster scales.

At lower temperatures, the isothermal and preheated models diverge.  
At $z=0$, $N(>kT)$ rises sharply below $kT=1$ keV in the preheated model.  This is due to the
nearly flat relation between $T$ and $M$ as shown in Figure \ref{M-T}:
halo masses between $10^{12.5}$ and $10^{14}$ at $z=0$ have nearly the same temperature of $kT=0.8$ keV.
This is the mass range where the gas mass fraction is fixed at 
$\Omega_b/\Omega_m$, and $kT(R_h)>{1 \over 2}kT_h$, and we have discussed the cause
of the constant temperature in this regime, in \S\ref{sec-lt}.
The model therefore predicts an overabundance of clusters within a narrow temperature range
around $kT=0.8$ keV, relative to the isothermal model.  Below this
temperature, the temperature function of the isothermal and preheated models become
approximately parallel again, since the slopes of their respective $M-T$ relations are 
not very different.
The direction of evolution of the temperature function below $kT=1$ keV also changes to
positive evolution in the case of the preheated models.  This is because the ``constant
temperature'' regime occurs at a higher temperature at higher redshifts.  Therefore, the
sharp increase in the temperature function occurs at $kT=2$ keV at $z=1$, for example.

\subsubsection{The Luminosity Function}
The differential luminosity function is shown in Figure \ref{fig-Ldif}.  The data are taken from the
RDCS \cite{RDCS} and the EMSS \cite{EMSS,EMSS2}. 
We correct the data to our chosen cosmology, and correct the luminosities to bolometric
values.  For the models, we use the values
of $\sigma_8$ chosen to match the temperature function, namely $0.78$ for the preheated
models, and $0.88$ for the isothermal models.  

\begin{figure*}
\begin{center}
\leavevmode \epsfysize=8cm \epsfbox{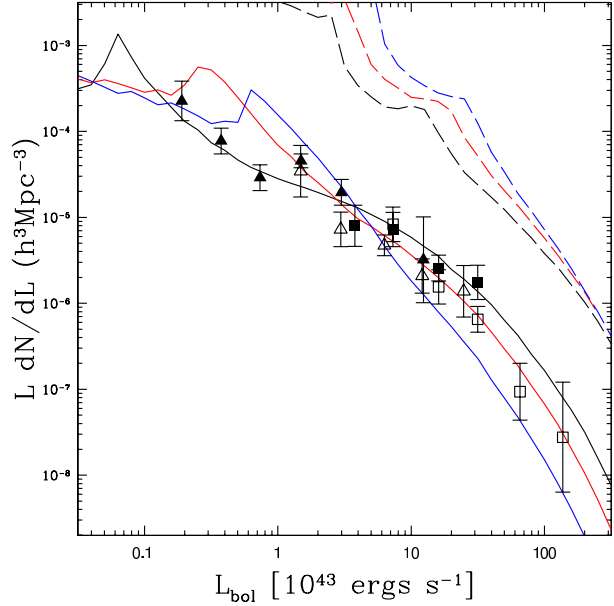}
\end{center}
\caption{The differential luminosity function.  Data are from the
RDCS ({\it triangles}) and the EMSS ({\it squares}), converted
to our cosmological parameters.  Solid symbols are the low redshift data
($z<0.2$); the open symbols are high redshift clusters ($0.3<z<0.6$).  
The {\it solid lines} are the preheated
models at $z=0$, $z=0.5$, and $z=1$, such that $N(>L)$ decreases with redshift at
high luminosities.  The {\it dashed lines} are isothermal models at the same
redshifts; evolution is strictly positive in these models.
\label{fig-Ldif}}
\end{figure*}

While the preheated model matches the $z=0$ data very well, the normalisation of the
isothermal model is much too high.  From the luminosity-temperature relation 
(Figure \ref{L-T})
it is clear that this should be so.  For a given temperature, the isothermal luminosities
are too high, even in the most massive clusters, where the gas is expected to be approximately
isothermal.  In particular, this is different from the result in Balogh, Babul, Patton (1999), 
where the isothermal luminosity function was not so discrepant with the data.  The reason 
is that, in the model of Balogh, Babul, Patton (1999), we artificially introduced a flat 
core in the isothermal potential, to prevent the luminosities from diverging.
As the size of this flat region is decreased, the luminosity of a given mass halo increases.  
In the current isothermal models there is no explicit need for such artificial structures 
since the potential we use does not lead to a divergent luminosity.  However, the 
potential is necessarily steeper in the centre, which leads to higher densities and 
higher luminosities.

In the preheated models, the kink in the $z=0$ luminosity function at $10^{42}$ ergs s$^{-1}$ is
due to a sharp change in slope of the $M-L$ relation, as shown in Figure \ref{fig-MLfig}.  
This corresponds to the point in
the models where the gas mass fraction is equal to $\Omega_b/\Omega_m$, and there
is a discontinuity in $dM/dL$.
In the isothermal models, the non-monotonic shape at low luminosity is due to
the corresponding very low temperatures (see Figure \ref{L-T}), where 
cooling is dominated by line emission
and the cooling function changes rapidly with temperature.  The shape of the luminosity
function in this temperature regime will be quite sensitive to the metallicity, for that
reason.
  
\begin{figure}
\begin{center}
\leavevmode \epsfysize=8cm \epsfbox{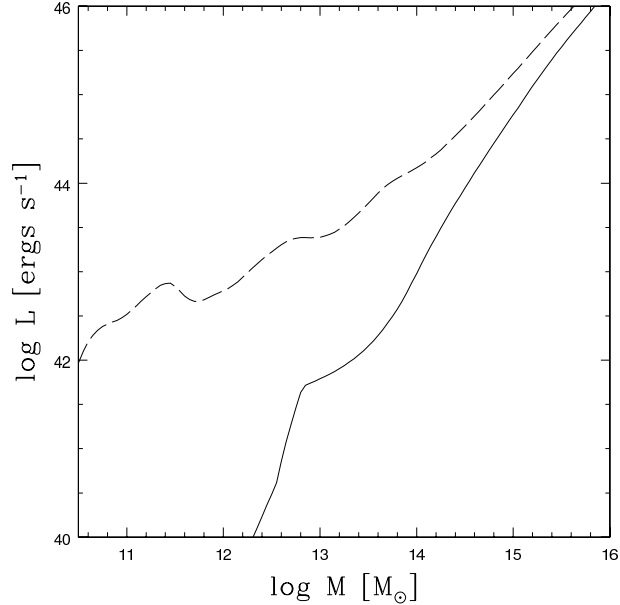}
\end{center}
\caption{The relationship between mass and luminosity for 
the isothermal ({\it dashed line}) and preheated ({\it solid line}) models.
The isothermal model is overluminous, relative to the preheated models, due to
the steep central potential.  This leads to the difference in luminosity
functions seen in Figure \ref{fig-Ldif}.
\label{fig-MLfig}}
\end{figure}

The preheated model predicts mild negative evolution of 
the luminosity function at bright luminosities.  This is in quite good agreement with
the data, which show such evolution for $L>10^{44}$ ergs s$^{-1}$, but no evolution at
lower luminosities.  In contrast, the isothermal models show weak evolution at high
luminosities, and stronger {\it positive} evolution at lower luminosities.  

\section{Discussion}\label{sec-discuss}

There has been considerable interest in comparing and contrasting the correlations
between the various  X-ray properities, and between the X-ray and the optical properties,
of groups and clusters.   Although, at present there is no clear consensus on what the 
trends are indicating (Mulchaey 2000), studies of these kind can potentially provide 
considerable insight into the nature of the ICM in groups and clusters, and into the 
dominant mechanisms underlying their formation and evolution.   While there is no
doubt that to a fair extent the lack of consensus stems from the fact that the 
X-ray and optical properties of groups as not well determined --- as discussed by 
Mulchaey (2000), the X-ray and optical  properties of poor groups are subject to 
statistical and systematic uncertainties caused by small number statistics in both group members and X-ray 
photons --- the fact that the discussions of the correlations have tended to 
focus on very different aspects also adds to the current state of affairs.

\begin{figure}
\begin{center}
\leavevmode \epsfysize=8cm \epsfbox{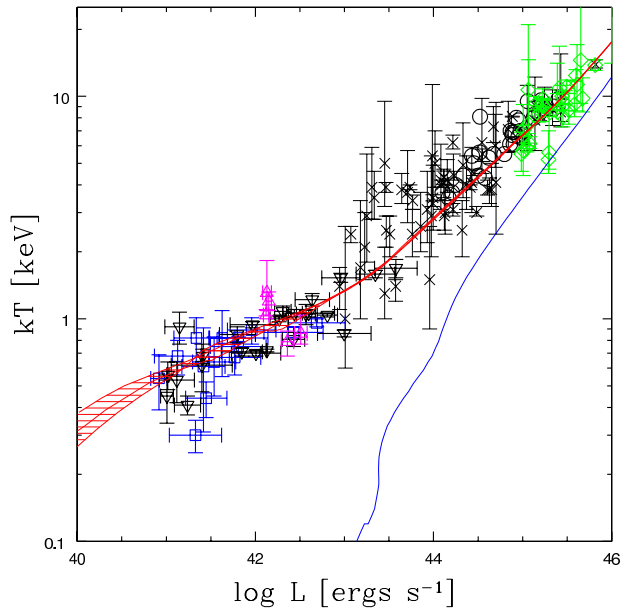}
\end{center}
\caption{The $z=0$ luminosity--temperature relation for the preheated model
({\it hatched region}) and the isothermal model ({\it solid line}), in which
the luminosity and temperature are computed by 
integrating the X-ray emissions out to the typical radius characterizing the 
spatial extent of the observed X-ray emissions from systems of a given 
temperature.  The data are the same as in Figure \ref{L-T}; however the
luminosities of the low temperature systems have not been corrected for
emission outside the observed radius (since we now take account of this
in the models).
In constrast with the results presented in Figure \ref{L-T}, the entropy
constant of the ``preheated'' model that best fits the data is $K_\circ=0.35 K_{\rm 34}$ 
or $kTn_e^{-2/3}\approx 330\kev\cm^2$.
}\label{mulchaey-lt}
\end{figure}

In an effort to help establish a common basis for discussing the commonalities
and differences between the X-ray and optical properties of groups and clusters,
we consider  four questions:\hfill\break

\medskip
\noindent
{\bf 1) Are the X-ray and optical trends of the clusters and groups accounted for by
the standard isothermal model for their formation and evolution?}  Figures 
\ref{L-T} and \ref{L-Sig}, for example, facilitate a comparison of the model
predictions and the observed results.  These figures clearly indicate that the
isothermal model fails to account for the observations.  However, there are both
observational and theoretical caveats to consider.
In computing
the X-ray luminosities of our theoretical haloes, we integrated the emissivity
over the entire halo and specifically, out to the virial radius in the plane
of the sky.  While the X-ray emission from 
rich clusters is indeed detected out to the cluster periphery,
Mulchaey (2000), Roussel \etal (2000) and \cite{HP00} have pointed out that only the
central regions of groups are detected, as the low surface brightness regions
are overwhelmed by the X-ray background.  Consequently, 
only a small fraction of the gas and thus, the X-ray luminosity, is directly 
detected in low temperature systems.  

There are two strategies for working around the problem of differing X-ray extents
of emissions from groups and clusters.  One option is to attempt to correct the 
observations for the ``missing flux'', in effect estimating the luminosity
that would have been detected had the spatial extent of the observed X-ray emissions
extended out to the virial radius in the plane of the sky, and then compare the 
corrected luminosity to theoretical results.  This is the strategy that has been 
adopted, for example, by Mulchaey (2000) in the construction of his Figure 3 and 
by Helsdon and Ponman (2000). Such corrections, however, require either an 
assumption of a model for the properties and the spatial distribution of the 
gas outside the observed region or the adoption of some scheme (\eg~the beta
model) to extrapolate the distribution of the X-ray surface brightness distribution 
in groups out to their virial radii.  Either approach is prone to gross systematic
uncertainty.
A better approach for carrying  out ``fair comparisons'' between observations and theory 
is to subject the theoretical results to similar observational and instrumental biases 
and limitations as the observations. In essence, the goal is to create mock observations that are
as realistic as possible and analyze these observations in exactly the same way 
as the real observations.  We are in the process of carrying out a detailed
analyses of this kind (Poole et al, in preparation) but, as a prelude,  we
have used the relationship between the observed temperature of a system and its 
projected radial extent of the detected X-ray emissions, derived from the observations
and kindly provided to us by Mulchaey (2000, private communications), to construct
theoretical luminosity-temperature curves that attempt to take into account, admittedly
very crudely,  the limited extent of detected X-ray emissions from low temperature systems.  
However, this has almost no effect on the isothermal models, as shown in Figure \ref{mulchaey-lt}.
This is because the temperature is fixed relative to the virial temperature and,
furthermore, the steep density profiles mean that the luminosities are dominated by 
emission at $R<0.1R_{\rm vir}$.  There is a larger effect on the preheated model,
and a lower entropy constant of $K_\circ=0.35 K_{\rm 34}$ 
or $kTn_e^{-2/3}\approx 330\kev\cm^2$ is required to fit the data.
This is a factor 
of 0.78 smaller than in the ``uncorrected'' case.  Comparing the 
original and truncated X-ray luminosities, we find a correction factor
of $\sim 2$ for systems with $kT < 1\kev$, which is in good agreement with
the correction factors adopted by Helsdon and Ponman (2000) in their
analyses.  For hotter systems, the correction factor drops sharply
to $\sim 1.3$ for systems with $kT \sim 2\kev$ and to $\sim 1.1$ by
$kT = 3\kev$.  We note also that the dependence on
formation epoch is greatly reduced in the low mass systems.
While we recognize that our efforts to incorporate observational biases in the
calculation of the theoretical curves are rather crude, the results do suggest
that the isothermal model does not describe the gas distribution in groups or 
clusters.

\medskip
\noindent
{\bf 2) Do the X-ray and optical correlations of groups an clusters scale similarly?}
This question is more difficult to address without actually going through the 
process of making mock but realistic  observations of groups exhibiting similar 
correlations.  We are in the process of doing so (Poole \etal in preparation).

At the moment, given the large uncertainties in the measurements of group properties 
and the fact that the  extent of the measured X-ray emissions from groups covers a small
fraction of their projected surface area, Figures \ref{L-T} and \ref{L-Sig}
do not rule out the possibility that 
the X-ray/optical correlations of groups and clusters scale in exactly the same way.  
The limited radial extent of the observations alone will cause departures from self-similar scaling on group 
scales resulting in steepening of the luminosity-temperature relationship (flattening
in our Figure \ref{L-T}).  Whether the observed steepening can be accounted for by
this effect is the one of the central issues that has yet to be resolved.  

However, even if cluster scaling relations  provide a suitable description of the 
observations across the entire range, from rich clusters to groups, one is still
faced with a puzzling conundrum: What is the theoretical basis for the scaling
relations?  What is the best model for describing the spatial, the dynamical and the
thermodynamical state of the hot intracluster medium in the haloes?  As discussed above, 
the isothermal model is certainly not it.

\medskip
\noindent
{\bf 3) Is there then  a theoretical model that can account for the observed trends of both
groups and clusters?}
Based on the analyses presented in this paper, we would argue that the ``preheated''
model is just such a model.  The ``preheated'' model appears to be able to account 
for most of the observed X-ray and optical correlations and  the more careful 
analyses of the kind presently underway will allow us to further strengthen the 
model's footing.  In the process, we also expect that some of the details, such 
as the optimal range for model parameters including the entropy constant $K_\circ$, may
change.

Much more importantly, there are some fundamental issues associated with the ``preheating''
 model that have yet to be resolved:  What is the cause of the preheating?  Does preheating 
affect only limited volumes of the universe, volumes that eventually collapse to form 
groups and clusters, or is it ubiquitous?  In the introduction, we briefly discussed some 
of these issues; however, there is considerably more work to be done.

\begin{figure}
\begin{center}
\leavevmode \epsfysize=8cm \epsfbox{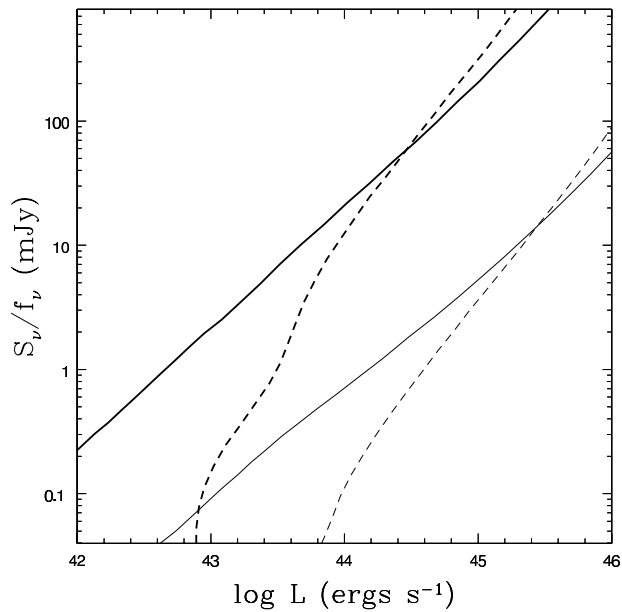}
\end{center}
\caption{Correlations between the integrated SZ flux and the X-ray
luminosity for clusters and groups placed at z=0.1 ({\it heavy
curves}) and z=1.0 {\it light curves}) for the preheated ({\it solid}) and 
isothermal ({\it dashed}) models. 
}\label{lsz-lx}
\end{figure}

\medskip
\noindent
{\bf 4) Are there tests that one can subject the preheated model to?}
The preheated model has a clear signature in the density and entropy
profiles, which can distinguish it from isothermal models.
With the high sensitivity and good resolution of
the {\it Newton} and {\it Chandra} telescopes, it is now possible
to observe these profiles directly (\eg~David \etal~2001). 
However, to make this comparison
between data and models fairly, it will
be necessary to create realistic mock observations from the models
and to ``reduce'' them in a way that is analagous with the real data,
to properly account for effects such as spatial resolution, projection,
and the limits of the energy bandpass (Poole \etal, in preparation).

The Sunyaev-Zel'dovich (SZ) effect  \cite{SZ,SZ2,R95,B99} 
offers another potential probe of the thermal state of the 
intracluster medium in groups and clusters.  Briefly,
cosmic microwave background (CMB) photons passing through
structures like galaxy clusters as they propagate towards us, 
will tend to get inverse Compton scattered by the hot electrons 
in the intracluster medium.  While conserving the number
of photons, the process does result in a preferential increase
in the energy of the photons and hence, a distortion in the 
CMB spectrum in the direction of the structure.

Although long recognized as a potentially powerful tool with 
which to study the intra- as well as intercluster/intergalactic 
medium, SZ measurements have proven to be very difficult.  
Advances in detector technology, observing techniques,
and the sheer perseverance of the people involved have,
however, have brought the field to the threshold of maturity.

The study of the SZ distortion offers an independent means 
(from the studies of X-ray emissions)  of  probing the thermal 
state and the spatial distribution of intracluster gas 
in groups and clusters.  In fact, SZ studies should prove to
be particularly revealing since the SZ flux density of a cluster, 
integrated over its face, is proportional to the total thermal
energy of the ICM \cite{B00}. 

In recognition of the above, we are carrying out a detailed study
of SZ effect as a test of the isentropic model and of the 
``preheating'' scenario (McCarthy \etal in preparation).  
As a prelude, we show in Figure
\ref{lsz-lx}, the plot of the ratio $S\nu/f\nu$ for groups
and clusters, in units of mJy, against their X-ray luminosity.
Here, $S\nu$ is the total SZ flux density from a cluster, found
by integrating the surface brightness over
the face of the group/cluster (\cf~Bartlett 2000):
\begin{equation}
\begin{split}
S_\nu&(x,M,z,z_{\rm form})=j_\nu(x)D_a^{-2}(z)\;\;\;\;\\
&\;\;\;\times \int{kT(M_{\rm form},z_{\rm form}) 
\over m_e c^2}n_e(M_{\rm form},z_{\rm form})\sigma_T\;\;dV,
\end{split}
\end{equation}
where the integral is over the entire virial volume of the cluster, $D_a$ 
is the angular diameter distance, $\sigma_T=0.6652 \times 10^{-24}$c
m$^{2}$ is the Thompson cross section,
and the function $j_\nu(x)$ describes the shape of the spectrum, as 
a function of the dimensionless frequency $x=h\nu/kT_\circ$.  
$T_\circ=2.78$ K is the temperature
of the unperturbed CMB spectrum, and $j_\nu(x)=2(kT_\circ)^3(hc)^{-2}f_\nu(x)$,
with
\begin{equation}
f_\nu(x)={x^4\mbox{e}^x \over \left(\mbox{e}^x-1\right)^2}\left[{x \over 
\mbox{tanh}(x/2)} -4\right].
\end{equation}
The plot also shows the predictions for the standard isothermal
model.  

Over four orders of magnitude in luminosity, the preheated model SZ flux scales
almost exactly proportionally with luminosity.  Even at high luminosities, the slope
is very different from that of the isothermal model, and this difference increases
toward lower luminosities.  We can therefore expect to obtain interesting constraints
from even the brightest SZ sources.

\section{Summary}\label{sec-conc}
We have constructed a physical model for the gas distribution in dark matter haloes, assuming
the gas has been preheated to a uniform but otherwise arbitrary entropy.  Apart from
this constraint, the critical assumptions are:
\begin{itemize}
\item Gas in virialised haloes is in pressure-supported hydrostatic equilibrium.
\item Gas is accreted at the Bondi rate, up to the maximum that can be accreted
gravitationally.
\item Some fraction, between 0 and 1, of the gas is accreted adiabatically, and
settles in the bottom of the potential to form an isentropic core.  The size of
this core is governed by a critical halo mass, below which we expect shocks to
be negligible.
\item Outside the isentropic core, the gas is shock heated.  We assume a linear
dependence of entropy on radius in this region, constrained to match the results
of numerical simulations.
\end{itemize}
In addition to the detailed construction of the models, we have shown a variety
of global scaling relations, between halo mass, velocity dispersion, and gas
temperature and luminosity.  We have made a detailed and critical comparison with data,
and shown that the model is
able to match the observed relations very well, over many orders of magnitude.
The minimum entropy required is
$K_\circ=0.45 K_{\rm 34}$ or $kTn_e^{-2/3}\approx 427\kev\cm^2$.  However, the
strongest constraints on the models come from low-temperature systems like
galaxy groups, and the observational biases in this data is troublesome enough
that we anticipate the level of this entropy minimum is yet to
be determined precisely.  

\section*{Acknowledgements}
First and foremost, we would like to thank Dr J. Raymond for providing us with
the latest version of his plasma routines.   We are indebted to
John Mulchaey, Richard Bower, Trevor Ponman, Neal Katz, Tom Quinn, David Spergel,
Ed Turner, Ian McCarthy and Gil Holder for many useful and relevant discussions 
during the course of this work.  MLB is supported by a PPARC rolling grant for
extragalactic astronomy and cosmology at Durham.  GBP gratefully acknowledges 
fellowship support from the University of Victoria.  AB gratefully acknowledges
the kind hospitality shown to him by the Institute for Theoretical Physics
(ITP) during the course of the Galaxy Formation workshop (January-April 2000)
where some of the work described in this paper was carried out.  AB also
acknowledges the hospitality of the University of Washington, and especially
T. Quinn, during his tenure there as Visiting Professor from May-August 2000.  
This research has been partly supported by the National Science Foundation 
Grant No.  PHYS94-07194 to ITP, NASA Astrophysics Theory Grant NAG5-4242 
to T. Quinn, as well as by an operating grant from the Natural Sciences and 
Engineering Research Council of Canada (NSERC).

\end{document}